\documentclass[aps,prb,reprint,superscriptaddress,showpacs,amsfonts]{revtex4-1}

\usepackage{epsfig}
\usepackage{amsmath}
\usepackage{amssymb}
\usepackage{color}
\usepackage{bm}

\newcommand{\beq}{\begin{equation}}
\newcommand{\eeq}{\end{equation}}
\newcommand{\beqarray}{\begin{eqnarray}}
\newcommand{\eeqarray}{\end{eqnarray}}

\newcommand{\eq}[1]{Eq.~(\ref{#1})} 
\newcommand{\fig}[1]{Fig.~\ref{#1}} 

\allowdisplaybreaks

\begin{document}

\title{Spin Josephson effect with a single superconductor}
\author{P. M. R. Brydon}
\email{brydon@theory.phy.tu-dresden.de}
\affiliation{Institut f\"{u}r Theoretische Physik, Technische Universit\"{a}t  
  Dresden, 01062 Dresden, Germany}
\author{Yasuhiro Asano}
\affiliation{Department of Applied Physics, Hokkaido University,
    Sapporo, 060-8628, Japan}  
\author{Carsten Timm}
\email{carsten.timm@tu-dresden.de}
\affiliation{Institut f\"{u}r Theoretische Physik, Technische Universit\"{a}t  
  Dresden, 01062 Dresden, Germany}

\date{\today}

\begin{abstract}
A thin ferromagnetic layer on a bulk equal-spin-pairing triplet superconductor
is shown to mediate a Josephson coupling between the spin $\uparrow$ and
$\downarrow$ condensates of the superconductor. By deriving analytic
expressions for the bound states at the triplet superconductor-ferromagnet
interface, we show that this spin Josephson effect establishes an effective
anisotropy axis in the ferromagnetic layer. The associated Josephson spin
current is predicted to cause a measurable precession of the magnetization
about the vector order parameter of the triplet superconductor. 
\end{abstract}

\pacs{74.50.+r, 74.20.Rp}

\maketitle 

\emph{Introduction.} 
The complex relationship between superconductivity and magnetism has motivated
an enormous effort to understand the properties of heterostructure interfaces
between ferromagnets (FMs) and spin-singlet superconductors
(SSCs).~\cite{SCFMreviews} A remarkable feature of such devices is the
existence of proximity-induced spin-triplet superconducting correlations due
to the exchange splitting in the FM, which are responsible for the anomalous
dynamics of the barrier magnetization in an SSC-FM-SSC Josephson
junction.~\cite{JJmagdyn} Due to the intimate connection between
ferromagnetism and triplet superconductivity, it is natural to consider what
results if the SSC were replaced by a triplet superconductor (TSC). This
question is of fundamental interest, as the intrinsic spin structure of the
Cooper pairs in a TSC allows us to anticipate an unconventional and unique
interplay with magnetism, which may be unambiguous signatures of the triplet
pairing state in proposed TSCs such as LiFeAs.~\cite{LiFeAs} For example, bulk
spin supercurrents are known to be possible in 
TSCs,~\cite{Asano,Linder2007,TFT,Brydon2009} and some proposals for their
realization require FM elements. The response of the FM component of the
device to the spin supercurrent, however, has yet to be investigated. The
recent fabrication of superconducting thin films of the suspected TSC
Sr$_2$RuO$_4$ is an important step towards the creation of TSC-FM
heterostructures,~\cite{MacMae2003,Krockenberger2010} and so a deeper
understanding of the physics of TSC-FM interfaces is timely.  

In this paper we show that a thin FM layer on a bulk equal-spin-pairing TSC
produces a spin Josephson effect by coupling the spin $\uparrow$ and
$\downarrow$ Cooper pair condensates. The physical mechanism is the
spin-dependent phase shift acquired by a Cooper pair undergoing spin-flip
reflection at the FM interface, which acts analogously to the phase difference
in a Josephson junction. Making only the assumption of spatially constant
order parameters, we solve the Bogoliubov-de Gennes (BdG) equations for the
bound states at the TSC-FM interface. Using this we calculate the free energy
of the interface, revealing that the spin Josephson effect creates an
effective hard or soft axis within the FM layer, depending upon the orbital
structure of the TSC gap. Finally, we obtain a general expression for the spin  
current, thereby showing that it exerts a measurable torque on the FM
moment. We propose this effect as a test for triplet pairing. 

\begin{figure}
\includegraphics[width=0.8\columnwidth]{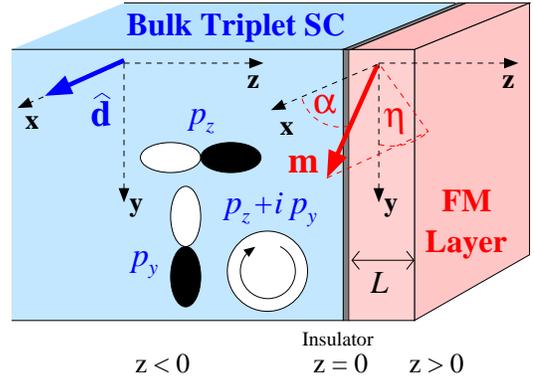}
\caption{\label{device}(Color online) Schematic diagram of the device
  studied here. The three different choices for the TSC orbital are shown: for
the $p_z$ and $p_y$ cases, the white and black lobes indicate opposite signs;
the arrow in the $p_z+ip_y$ shows the direction of increasing phase.} 
\end{figure}

\emph{Model system.} We consider a FM layer of width $L$
on a bulk TSC, separated by an atomically thin insulating layer,
see~\fig{device}. The BdG equation for the quasiparticle states with
energy $E$ is 
\beq
\left(\begin{array}{cc}
\hat{H}_{0}({\bf r}) & \hat{\Delta}({\bf r}) \\
\hat{\Delta}^{\dagger}({\bf r}) & -\hat{H}^{T}_{0}({\bf r})
\end{array}\right)\Psi({\bf r}) = E\Psi({\bf r})\; , \label{eq:BdG}
\eeq
where the caret indicates a $2\times2$ matrix in spin space. The wavefunction
$\Psi({\bf r})$ is only non-zero for $z<L$.
 The
non-interacting Hamiltonian $\hat{H}_{0}({\bf r})$ is 
\beq
\hat{H}_{0}({\bf r}) =
\left[\left(-\frac{\hbar^2\pmb{\nabla}^2}{2m}-\mu\right) + 
  U\delta(z) \right]\hat{\mathbf{1}} - \mu_{B}\hat{\pmb{\sigma}}\cdot{\bf
  {H}}_{\text{ex}}\Theta(z)\; . \label{eq:Ho} 
\eeq
Here $\mu_B$ is the Bohr magneton, and we make the simplifying but
nonessential assumptions that the radius of the Fermi surface $k_{F}$ and the
effective mass $m$ in the normal state of the TSC and the FM are the same.
The insulating layer is modeled as a $\delta$-function of strength $U$. The
last term in~\eq{eq:Ho} is the energy due to the exchange field ${\bf
  H}_{\text{ex}}$ in the FM. For an incompletely polarized FM, the
magnetization ${\bf m} = |{\bf m}|[\cos(\alpha){\bf e}_x +
  \sin(\alpha)\cos(\eta){\bf e}_y + \sin(\alpha)\sin(\eta){\bf e}_z]$ is
related to ${\bf H}_{\text{ex}}$ by Luttinger's theorem. We assume a
two-dimensional system, so that the majority-spin (parallel to ${\bf
  H}_{\text{ex}}$, $s=+$) and minority-spin (antiparallel to ${\bf
  H}_{\text{ex}}$, $s=-$) Fermi surfaces have radius $k_{F,s} =
\sqrt{(1+s\lambda)}k_{F}$, where $\lambda = \mu_{B}|{\bf
  H}_{\text{ex}}|/\mu<1$. This gives $|{\bf 
  m}|=\mu_{B}^2m|{\bf H}_{\text{ex}}|/\pi\hbar^2$. For a half-metallic FM
there is a single Fermi surface of radius $\sqrt{2}k_{F}$, and the exchange 
splitting is fixed by the details of the system. 

The gap matrix is $\hat{\Delta}({\bf r}) =
\Theta(-z)i[\hat{\sigma}\cdot{\bf d}]\hat{\sigma}^{y}$ where ${\bf d}=
\widetilde{\Delta}\hat{\bf d}$ is the vector order parameter, assumed
constant throughout the TSC. $\widetilde{\Delta}$ is an operator, which for
Cooper pairs in a relative $p$-wave orbital state has the real space form
$\widetilde{\Delta}=-i\Delta(T){\bf n}\cdot{\pmb{\nabla}}/k_F$. The gap
magnitude $\Delta(T)$ is assumed to have weak-coupling temperature
dependence. The unit vector ${\bf n}$ defines the orbital state: ${\bf n} =
{\bf e}_z$ for $p_z$-wave; ${\bf n} = {\bf e}_y$ for $p_y$-wave; and ${\bf n}
= {\bf e}_z+i{\bf e}_y$ for $(p_z+ip_y)$-wave. The last choice is of greatest
relevance to Sr$_2$RuO$_4$ and LiFeAs,~\cite{MacMae2003,LiFeAs} while the
others have been proposed for (TMTSF)$_2$X (X = PF$_6$,
ClO$_4$).~\cite{Lebed2000} In the following it is convenient to express the
gap in terms of the Fourier transform of $\widetilde{\Delta}$, which is
written $\Delta_{\bf k}= \Delta(T){\bf n}\cdot{\bf k}/k_F$. We fix $\hat{\bf
  d}={\bf e}_x$ which defines a TSC where the Cooper pairs have $z$-component
of spin $S_{z}=\pm{\hbar}$ but the condensed part of the system is
unpolarized. Below we show that only the angle between $\hat{\bf d}$ and ${\bf
  m}$ is relevant for the spin Josephson effect, and so other orientations of
$\hat{\bf d}$ do not result in new physics.

\emph{Bound states.} We seek solutions of~\eq{eq:BdG} for states 
bound to the FM layer. The wavefunction of such a state has
the general form $\Psi({\bf k}_{\parallel};{\bf r}) = \Psi_{\text{TSC}}({\bf
  k}_{\parallel};{\bf r})\Theta(-z) + \Psi_{\text{FM}}({\bf
  k}_{\parallel};{\bf r})\Theta(z)$ and satisfies
$\lim_{z\rightarrow-\infty}\Psi({\bf k}_{\parallel};{\bf r}) = 0$ and 
$\Psi({\bf k}_{\parallel};{\bf r})|_{z=L}=0$. The momentum component parallel
to the interface, ${\bf k}_\parallel$, is a good quantum number due to
translational invariance. Solving the Andreev equations in the
TSC,~\cite{Hu1994,KashTana2000} we make the ansatz 
\beq
\Psi_{\text{TSC}}({\bf k}_{\parallel};{\bf r}) =
\sum_{\sigma=\uparrow,\downarrow}\left[a_{1,\sigma}\Psi_{\sigma}({\bf k}_1;{\bf r}) +
{a}_{2,\sigma}\Psi_{\sigma}({\bf k}_2;{\bf r})\right]\; , \label{eq:TSCwf}
\eeq
where the spinors are given by
\beqarray
\Psi_{\uparrow}({\bf k};{\bf r})
 &=& \left(1,\; 0, \; \gamma({\bf k}), \;
0\right)^{T}e^{i{\bf k}\cdot{\bf r}}e^{\kappa_{\bf k}z}\; , \\ 
\Psi_{\downarrow}({\bf k};{\bf r}) 
 &=& \left(0, \; 1, \; 0, \;  -\gamma({\bf k})
\right)^{T}e^{i{\bf k}\cdot{\bf r}}e^{\kappa_{\bf k}z}\; ,
\eeqarray
with $\gamma({\bf k}) = -[E + i\mbox{sgn}(k_z)\sqrt{|\Delta_{\bf k}|^2 -
    E^2}]/\Delta_{\bf k}$ and $\kappa_{\bf k} =
(m/\hbar^2|k_z|)\sqrt{|\Delta_{\bf k}|^2 - E^2}$. The wavevectors appearing
in~\eq{eq:TSCwf} are defined by ${\bf k}_1=({\bf k}_{\parallel},k_z)$, ${\bf
  k}_2=({\bf k}_{\parallel},-k_z)$. Note that $|\Delta_{{\bf
    k}_1}|=|\Delta_{{\bf k}_2}|\equiv|\Delta_{{\bf k}_\parallel}|$ for the
orbital symmetries considered here.  

Depending upon the value of ${\bf k}_{\parallel}$, we have either propagating
or evanescent solutions in the FM layer. In the case when there are
propagating solutions in both spin channels we have 
\beqarray
\Psi_{\text{FM}}({\bf k}_{\parallel};{\bf r}) & = &
\sum_{s=\pm}\left\{b_{e,s}\sin(k_{e,s}[L-z])e^{i{\bf
    k}_{\parallel}\cdot{\bf r}}\Phi_{e,s}\right. \notag \\
&& + \left.b_{h,s}\sin(k_{h,s}[L-z])e^{i{\bf
    k}_{\parallel}\cdot{\bf r}}\Phi_{h,s}\right\} \label{eq:FMwf}
\eeqarray
where the electron and hole spinors are defined by
\beq
\Phi_{e,s} = \left(w_s, \; x_s, \; 0, \; 0 \right)^{T}\, , \quad
\Phi_{h,s}
=\left(0, \; 0, \; w_s^\ast, \; x_s \right)^{T}\, ,
\eeq
with $w_s = s({\cos\alpha - i\sin\alpha\cos\eta})/{\sqrt{1 -
    s\sin\alpha\sin\eta}}$, $x_s = \sqrt{1 - s\sin\alpha\sin\eta}$, and the 
wavevector $k_{e(h),s}$ for electrons (holes) is $k_{e(h),s} = [k_{F}^2(1 +
  s\lambda)- |{\bf k}_\parallel|^2 +(-)2m E/\hbar^2]^{1/2}$. If the radicand
is negative, only evanescent solutions are possible; in this case we replace
$k_{e(h),s}\rightarrow i\kappa_{e(h),s}$ where $\kappa_{e(h),s}$ is the
inverse decay length. 

The coefficients in~Eq.s~(\ref{eq:TSCwf}) and~(\ref{eq:FMwf}) are chosen so
that at the TSC-FM interface the wavefunction is continuous   
$\Psi({\bf k}_{\parallel};{\bf r})|_{z=0^-} = \Psi({\bf k}_{\parallel};{\bf
  r})|_{z=0^+}$, and its derivative obeys ${\partial_z}\Psi({\bf
  k}_{\parallel};{\bf r})|_{z=0^+}-{\partial_z}\Psi({\bf
  k}_{\parallel};{\bf r})|_{z=0^-} = 2Z\Psi({\bf k}_{\parallel};{\bf
  r})|_{z=0^+}$ where $Z = mU/\hbar^2$. The values of $E$ for which the
determinant of the resulting system of equations vanishes define the
bound-state energies. Explicit expressions for the bound-state energies can be
found when the $E$-dependence of the wavevectors is neglected,
i.e. $k_{e,s}\approx k_{h,s}\approx k_s$. This approximation is valid for a
thin FM layer such that $(k_{e,s} -  k_{h,s})L \approx 2EL/\hbar v_{F,s} \ll
1$.~\cite{KashTana2000,longFM} For a weakly to moderately polarized FM layer
we have Fermi velocities $v_{F,+}\approx v_{F,-} \sim 10^{6}$m$s^{-1}$, and so
for $E \leq \max\{|\Delta_{\bf k}|\} \sim 0.1$meV ($T_{c}\sim1$K) we require
thin layers less than about $100$ unit cells thick. In this limit we obtain
the non-degenerate bound states:    
\begin{subequations}
\label{eq:boundstates}
\begin{xalignat}{1}
E_{\pm,{\bf k}_\parallel} & = \; \pm|\Delta_{{\bf k}_\parallel}|\sqrt{D_{{\bf
      k}_\parallel}}|\cos\alpha|,  \phantom{::}\qquad\quad p_z\text{-wave}   \label{eq:pzstate}\\
E_{\pm,{\bf k}_\parallel} & = \; \pm|\Delta_{{\bf k}_\parallel}|\sqrt{1 - D_{{\bf
      k}_\parallel}\cos^{2}\alpha},\qquad p_y\text{-wave} \label{eq:pystate} \\
E_{\pm,{\bf k}_\parallel} & = \;  -|\Delta_{{\bf k}_\parallel}|\left[\sqrt{1-
    D_{{\bf k}_\parallel}\cos^{2}\alpha}\frac{k_y}{k_F}\right.  \notag \\
& \phantom{=}\;\;\left. \pm \sqrt{D_{{\bf k}_\parallel}}\cos\alpha\frac{k_z}{k_F} \right]. \phantom{:}\quad (p_z+ip_y)\text{-wave} \label{eq:chiralstate}
\end{xalignat}
\end{subequations}
Here we have
\beqarray
D_{{\bf k}_{\parallel}} & = &
4\left[\sum_{s=\pm}s\widetilde{k}_{s}\cos(k_sL)\sin(k_{{-s}}L)\right]^2 \notag
\\
&& \times\prod_{s=\pm}\left[1 + 4\widetilde{Z}^2 +
  \widetilde{k}_{s}^2 + 4\widetilde{k}_{s}\widetilde{Z}\sin(2k_{s}L)\right.\notag \\
&& \left. + (\widetilde{k}_{s}^2 -
  4\widetilde{Z}^2 - 1)\cos(2k_{s}L)\right]^{-1}
\eeqarray
where $\widetilde{A} = A/\sqrt{k_{F}^2-|{\bf k}_{\parallel}|^2}$
($A=k_{s},Z$). 

The bound-state energies~\eq{eq:boundstates} are a central result of our
paper. They originate due to multiple Andreev reflections within the thin FM
layer, which Josephson-couple the $S_{z}=\pm\hbar$ condensates in the TSC. The
same physical mechanism is responsible for the formation of Andreev bound
states (ABSs) at the tunneling barrier in a Josephson
junction.~\cite{BeenHout1992,ABSTriplet,KashTana2000} Remarkably, the
states~\eq{eq:boundstates} are identical to the (spin degenerate) ABSs in a
short Josephson junction of transparency $D_{{\bf k}_\parallel}$ between
$p_z$-wave TSCs with phase difference $\Delta\phi=2\alpha$ [\eq{eq:pzstate}],
between $p_y$-wave TSCs with $\Delta\phi=\pi + 2\alpha$ [\eq{eq:pystate}], and
between a $(p_z+ip_y)$-wave and $(p_z-ip_y)$-wave TSC with
$\Delta\phi=2\alpha$ [\eq{eq:chiralstate}]. Since the form of the bound-state
energies is fixed by the bulk pairing symmetry, our results should be robust
to a self-consistent calculation of the gap.~\cite{KashTana2000}

The spin Josephson coupling can also be understood at a more fundamental
level: in the BdG Hamiltonian the $S_{z}=\pm\hbar$ condensates in the TSC are
independent of one another. Tunneling of a Cooper pair between the two
condensates is made possible by the FM layer, where the coupling to the FM
moment allows an incident Cooper pair with spin $\sigma\hbar$ to be reflected
with spin $-\sigma\hbar$. As a result of this process, the Cooper pair
acquires a phase shift $\Delta\theta_{{\bf k}_\parallel} + \pi
-2\sigma\alpha$. The last terms are due to the spin flip itself, and are
primarily responsible for driving the spin current.
$\Delta\theta_{{\bf k}_\parallel} =  \arg\{\Delta_{{\bf k}_2}\} -
\arg\{\Delta_{{\bf k}_1}\}$ is the phase shift due to the orbital structure of
the TSC: for $p_z$-wave orbitals we have $\Delta\theta_{{\bf
    k}_\parallel}=\pi$ for all ${\bf k}_\parallel$; in the $p_y$-wave case we
have $\Delta\theta_{{\bf k}_\parallel}=0$; and the superposition of these two
orbitals in the $(p_z+ip_y)$-wave TSC gives $\Delta\theta_{{\bf k}_\parallel}
= \pi -2\arccos(\sqrt{1-|{\bf k}_\parallel|^2/k_F^2})$. $\Delta\theta_{{\bf
    k}_{\parallel}}$ accounts for the $\pi$ phase difference between the bound
states in the $p_z$ and $p_y$ cases, and the apparent sign reversal of the
$p_y$ component in the $p_z+ip_y$ bound states. 

We note that unlike the standard Josephson effect, where the phase difference
between the two superconductors drives the supercurrent, the spin Josephson
effect is due to phase shifts picked up \emph{during} the tunneling process
itself. The phase difference between the spin-up and spin-down condensates is
fixed by the orientation of $\hat{\bf d}$, which is unaffected by the FM
layer.  
 
\begin{figure}
\includegraphics[width=\columnwidth]{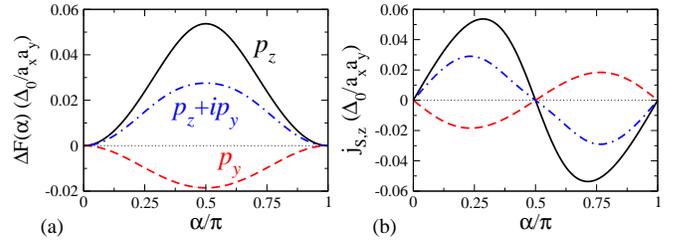}
\caption{\label{freeE+sc}(Color online) (a) The free energy difference
    $\Delta F(\alpha)$ and (b) the $z$-component of
    the spin current $j_{S,z}$ per interface unit cell area as a function of
    $\alpha$ for the three choices 
    of orbital wavefunction. We take $\lambda=0.05$, $T=0.4T_{c}$, $\eta=0$,
    $L=10a_{z}$ and $Z=1$. $\Delta_0$ is the $T=0$ gap magnitude.}
\end{figure}

\emph{Free energy.} The assumption of spatially constant order parameters
allows us to write the free energy due to the spin Josephson coupling
in the TSC-FM device in terms of the bound states~\cite{BeenHout1992}
\beq
F = -\frac{1}{2}k_{B}T\sum_{n=\pm}\sum_{{\bf
    k}_{\parallel}}\ln\left[2\cosh(\beta E_{n,{\bf k}_{\parallel}}/2)\right]
+ F_{0}\, , \label{eq:intF}
\eeq
where $F_0$ is independent of $\alpha$ and includes the interaction energy
in both the TSC and the FM, as well as the contribution from continuum
states. We plot the free energy difference
$\Delta{F}(\alpha)=F(\alpha)-F(\alpha=0)$ per interface-unit-cell area
$a_xa_y$ as a function of $\alpha$ in~\fig{freeE+sc}(a). As can be seen, $F$
takes a minimum as a function of the angle $\alpha$: regarding $\hat{\bf
  d}$ as a fixed property of the bulk TSC, the spin Josephson coupling
therefore establishes a preferred orientation for the magnetization of the FM
layer. There is a direct analogy to a short Josephson junction, where the free
energy due to the Josephson coupling takes a minimum as a function of
$\Delta\phi$.~\cite{BeenHout1992} 

In the $p_z$-wave ($p_y$-wave) case, the formally equivalent Josephson
junction with $\Delta\phi=2\alpha$ ($\Delta\phi=\pi-2\alpha$) has
time-reversal symmetry and so the Josephson free energy is minimized at
$\Delta\phi=0$. This implies that in the TSC-FM device the free energy
\emph{always} has a minimum at $\alpha=0$ ($\alpha= \pi/2$). The broken
time-reversal symmetry in the $(p_z+ip_y)$-wave case, however, means that the
stable value of $\alpha$ is determined by the details of $D_{{\bf
    k}_\parallel}$. Specifically, the $p_z$ and $p_y$ components of the gap
favor minima at different values of $\alpha$: if $D_{{\bf k}_\parallel}$ is
peaked near $|{\bf k}_{\parallel}| = k_F\, (0)$, the $p_y$-component
($p_z$-component) dominates and the configuration with $\alpha=\pi/2$ ($0$) is
stable; for more complicated $D_{{\bf k}_{\parallel}}$, the competition
between the gap components may stabilize the system at $\alpha\neq 0,\,
\pi/2$. For weak magnetization strengths, the free energy minimum is usually
located at $\alpha=0$. 

From~\fig{freeE+sc}(a) we see that to excellent approximation $F\propto\cos
2\alpha$. In writing an effective free energy for the FM layer, we can
therefore account for the spin Josephson effect by including a term $F_J =
f_{s}(\hat{\bf d}\cdot{\bf m})^2$, i.e. for $f_s<0$ ($f_s>0$), $\hat{\bf d}$
defines an effective easy (hard) axis in the FM layer. Since the sign of $f_s$
is determined by the \emph{orbital} state of the TSC, this reveals a novel
type of spin-orbit coupling between the TSC and the FM. 

\emph{Spin current and magnetization dynamics.} The zero-bias charge current
$I_J$ in a Josephson junction is given by $I_J =
(2e/\hbar)\partial{F}/\partial\Delta\phi$. We now show that in our device
there is a spontaneous spin current which can be similarly expressed as a
derivative of $F$ with respect to~$\alpha$. 

Our starting point is the continuity equation for the spin
\beq
{\bf J}_{s} = 
\frac{\hbar}{g\mu_B}\frac{d}{dt}{\bf M} =
{\bf M}\times\frac{\partial F}{\partial{\bf M}}\, , \label{eq:LLG}
\eeq
where $g$ is the gyromagnetic ratio and ${\bf M}$ is the total moment of the
FM layer. The vector notation for the spin current ${\bf J}_s$ refers
\emph{only} to the polarization; the direction of the spin current is normal
to the interface. We omit gradient terms in~\eq{eq:LLG} as the FM layer is
considered to be thin compared to the coherence length of the Cooper
pairs. The free energy in our problem has the form $ F = F(|{\bf M}|,\hat{\bf  
  d}\cdot\hat{\bf M})$, allowing us to write
\beqarray
{\bf J}_s & = & {\bf
  M}\times \left(\frac{\partial{|{\bf M}|}}{\partial {\bf
    M}}\frac{\partial{F}}{\partial{|\bf M}|} + \frac{\partial{\hat{\bf
      d}}\cdot\hat{\bf M}}{\partial{\bf M}}\frac{\partial{F}}{\partial{\hat{\bf d}}\cdot\hat{\bf M}} \right) \notag \\
& = & {\bf M}\times\left(\hat{\bf M}\frac{\partial{F}}{\partial{|\bf M}|} +
\frac{1}{|{\bf M}|}[\hat{\bf d} - (\hat{\bf d}\cdot\hat{\bf M})\hat{\bf
    M}]\frac{\partial{F}}{\partial \cos(\alpha)} \right) \notag \\ 
& = & \hat{\bf p}\frac{\partial F}{\partial \alpha}\, ,  \label{eq:dfda}
\eeqarray
where $\hat{\bf p} = \hat{\bf d}\times{\bf M}/|{\bf M}|\sin\alpha$ is a
unit vector which points in the same direction for all ${\bf M}$ lying in
a fixed plane containing $\hat{\bf d}$. Inserting~\eq{eq:intF}
into~\eq{eq:dfda} we obtain  
\beq
{\bf J}_{s} = -\hat{\bf p}\frac{1}{4}\sum_{n=\pm}\sum_{{\bf
    k}_\parallel}\frac{\partial E_{n,{\bf k}_\parallel}}{\partial
  \alpha}\tanh(\beta E_{n,{\bf k}_\parallel}/2)\,. \label{eq:scABS}
\eeq
This closely resembles the Beenakker-van Houten formula for the charge current
in a short Josephson junction.~\cite{BeenHout1992} It reveals that the spin
current in our device is due entirely to resonant tunneling between the two
spin condensates through the bound
states~\eq{eq:boundstates}. Equation~(\ref{eq:scABS}) gives identical results 
to the Furusaki-Tsukada technique,~\cite{FuruTsu1991,Asano,Brydon2009} which  
expresses the spin current in terms of the Andreev reflection coefficients. We
show the spin current as a function of $\alpha$ in~\fig{freeE+sc}(b). 

If the magnetization is prepared with $0<\alpha<\pi/2$,~\eq{eq:LLG} predicts
that the spin current will exert a torque on ${\bf M}$, causing it to precess
about $\hat{\bf d}$. Writing ${\bf M}=AL\mu_{B}pn\hat{\bf m}$ and ${\bf
  J}_s=A{\bf j}_{s}$, where $A$ is the area of the TSC-FM interface, and $p= 
(n_{+}-n_{-})/n$ is the polarization of the FM, we find the precession
frequency to be $\Omega_J = 2g\cos(\alpha)\text{max}\{|{\bf j}_s|\}/\hbar n p
L$. To estimate $\Omega_J$, we assume a weakly polarized FM, $\lambda = 0.05$,
with $n = 1$ electron per unit volume $v = a_xa_ya_z$, $\text{max}\{|{\bf
  j}_s|\} = 0.025\Delta_{0}/(a_xa_y)$ [see~\fig{freeE+sc}(b)], $L =10 a_z$,
and $T_c = 1$K. We hence find $\Omega_J = 15\cos(\alpha)\,$GHz, which is
measurable by ferromagnetic resonance (FMR) experiments. As the spin Josephson
effect cannot occur for a SSC, the observation of this precession would be
very strong evidence of a triplet pairing state, although the precession
effects due to multiple FM domains would have to be ruled
out.~\cite{Grein2009} Similarly, it is also necessary to examine the effect of
chiral domains of the gap in the candidate material
Sr$_2$RuO$_4$.~\cite{chiral} Gilbert damping and anisotropy effects in the
FM layer must also be included in a complete description of the magnetization
dynamics, but do not change the derivation of~\eq{eq:scABS}.

\emph{Conclusions.} In this paper we have demonstrated that the spin
structure of the Cooper pairs in a TSC permits the occurrence
of a spin Josephson effect without the need for a second superconductor. We
have proposed that a thin FM layer on a bulk TSC can realize this effect. In
turn, the spin Josephson coupling establishes an 
effective easy or hard axis in the FM layer, depending upon the orbital
symmetry of the TSC gap. Furthermore, the Josephson spin current causes the
magnetization to precess about the $\hat{\bf d}$ vector with a frequency that
is accessible to FMR, realizing a possible experimental signature of the
triplet state. 

The authors thank D. Manske, J. A. Sauls, and M. Sigrist for useful
discussions. Y. A. was supported by KAKENHI on Innovative Areas ``Topological
Quantum Phenomena" (No. 22103002) and KAKENHI (No. 22540355) from MEXT of
Japan.

\end{document}